\documentclass[twocolumn,superscriptaddress,longbibliography,aps,pra,preprintnumbers]{revtex4-1}

\usepackage{graphicx}
\usepackage{bm}
\usepackage{color}
\usepackage{epstopdf}
\usepackage{amsmath}
\usepackage{amssymb}
\usepackage{epstopdf}

\usepackage[urlcolor=blue,colorlinks=true,citecolor=blue,linkcolor=blue,pdfstartview={FitH},bookmarks=false]{hyperref}

\newcommand{\Tr}{\text{Tr}}

\graphicspath{{fig/}{./fig/}{.}}

\begin{document}

\title[Critical behaviour in one dimensional]{Critical behaviour in one dimension: unconventional pairing, phase separation, BEC-BCS crossover and magnetic Lifshitz transition}

\author{Andrzej Ptok}
\email[e-mail: ]{aptok@mmj.pl}
\affiliation{Institute of Physics, Maria Curie-Sk\l{}odowska University, Plac M. Sk\l{}odowskiej-Curie 1, PL-20031 Lublin, Poland}
\affiliation{Institute of Nuclear Physics, Polish Academy of Sciences, ul. E. Radzikowskiego 152, PL-31342 Krak\'{o}w, Poland}

\author{Agnieszka Cichy}
\email[e-mail: ]{agnieszkakujawa2311@gmail.com}
\affiliation{Institut f\"{u}r Physik, Johannes Gutenberg-Universit\"{a}t Mainz, Staudingerweg 9, D-55099 Mainz, Germany}

\author{Karen Rodr\'{i}guez}
\email[e-mail: ]{karem.c.rodriguez@correounivalle.edu.co}
\affiliation{Departamento de F\'{i}sica, Universidad del Valle, A.A. 25360, Cali, Colombia}
\affiliation{Centre for Bioinformatics and Photonics -- CiBioFi, Calle 13 No. 100-00, Edificio 320 No. 1069, Cali, Colombia}

\author{Konrad Jerzy Kapcia}
\email[e-mail: ]{konrad.kapcia@ifpan.edu.pl}
\affiliation{Institute of Physics, Polish Academy of Sciences, Aleja Lotnik\'{o}w 32/46, PL-02668 Warsaw, Poland}

\date{\today}

\begin{abstract}
We study the superconducting properties of population-imbalanced ultracold Fermi mixtures in one-dimensional (1D) optical lattices  that can be effectively described by the spin-imbalanced attractive Hubbard model (AHM) in the presence of a Zeeman magnetic field.
We use the mean-field theory approach to obtain the ground state phase diagrams including some unconventional superconducting phases such as the Fulde--Ferrell--Larkin--Ovchinnikov (FFLO) phase, and the $\eta$ phase (an extremal case of the FFLO phase), both for the case of a fixed chemical potential and for a fixed number of particles.
It allows to determine optimal regimes for the FFLO phase as well as $\eta$-pairing stability.
We also investigate the evolution from the weak coupling (BCS-like limit) to the strong coupling limit of tightly bound local pairs (BEC) with increasing attraction, at $T=0$.
Finally, the obtained results show that despite of the occurrence of the Lifshitz transition induced by an external magnetic field, the superconducting state can still exist in the system, at higher magnetic field values.
\end{abstract}

\pacs{PACS:}

\maketitle

\section{Introduction}

The immense development of experimental techniques in cold atomic Fermi gases in the last years has opened new avenues for research of strongly correlated systems in condensed matter physics and beyond.
The ability to control the interactions via Feshbach resonances~\cite{fedichev.kogan.96} sets new perspectives for experimental realization and study of many different unconventional systems, such as spin-polarized superfluidity (with population imbalance), superconductivity with nontrivial Cooper pairing, Bose-Fermi mixtures or mixtures of fermions with unequal masses~\cite{zwierlein.schirotzek.06, partridge.li.06,zwierlein.ketterle.06,georgescu.ashhab.14}.

There are indications that the properties of unconventional superconductors place them between two regimes: BCS and BEC~\cite{micnas.ranninger.90,
bourdel.khaykovich.04,kapcia.robaszkiewicz.12,cichy.micnas.14,kapcia.14}.
The evolution from the weak attraction (BCS-like) to the strong attraction (BEC-like) limit takes place when the interaction is increased or the particle concentration is decreased at moderate fixed attraction.
According to the Leggett criterion~\cite{leggett.80}, the Bose regime begins when the chemical potential $\mu$ drops below the lower band edge.
The possibility to control population imbalance has motivated attempts to understand the BCS-BEC crossover phase diagrams in the presence of spin polarization~\cite{wolak.gremaud.12,cichy.micnas.14}.

Currently, the unconventional superconductivity with a non-trivial Cooper pairing lays down one of the most important directions of studies in the theory of condensed matter~\cite{matsuda.shimahara.10} and ultracold quantum gases~\cite{hu.xiaji.06,dutta.gajda.15}.
In the presence of a Zeeman magnetic field, the densities of states are different for the particles with spin down and spin up.
In the case of ultracold Fermi gases, paramagnetic effects are introduced artificially by population imbalance producing a mismatch between the Fermi surfaces.
At strong imbalance, in the weak coupling regime, superfluidity is destroyed and undergoes a first-order phase transition to the polarized normal state at a universal critical magnetic field $h_{P} = \Delta_0 / \sqrt{2} \approx 0.707 \Delta_0$.
The latter is called the Chandrasekhar-Clogston (CC) limit or Pauli limit~\cite{chandrasekhar.62, clogston.62}, where $\Delta_0$ is the gap at zero temperature in the absence of external field.
Rather recently, a behaviour in accordance with the CC limit has been observed in population imbalanced atomic Fermi gases~\cite{zwierlein.schunck.06,cichy.cichy.15}.

In the weak coupling limit, at a large difference in the occupation number (or at a strong magnetic field), states with nontrivial Cooper pairing can exist.
An example of such pairing is the formation of Cooper pairs across the spin-split Fermi surface with non-zero total momentum (${\bf k} \uparrow$, $-{\bf k}+{\bf Q} \downarrow$), leading to the so-called Fulde--Ferrell--Larkin--Ovchinnikov~\cite{FF,LO} (FFLO) state.
Solid-state experiments typically involve highly anisotropic materials -- made up either of weakly coupled two-dimensional (2D) planes or 1D wires~\cite{matsuda.shimahara.10}.
The potential candidates for finding the FFLO phase are heavy fermions~\cite{bianchi.movshovich.02,radovan.2003,bianchi.movshovich.03,kenzelmann.2008,matsuda.shimahara.10}, organic~\cite{uji.2006,lortz.wang.07,mayaffre.kramer.14} or iron-based superconductors~\cite{zocco.grube.13,ptok.15,januszewski.ptok.15}.
These systems are characterized by a discontinuous phase transition from the superconducting to the normal state in the regime of low temperatures.
However, it is still unclear in which range of parameters the FFLO phase is stable.
Moreover, the observation of this type of superconductivity is very difficult because of the very strong destructive influence of the orbital (diamagnetic) effect.

For instance, some calculations indicate that if a FFLO phase exists in 3D trapped gases, it will occupy a very small volume in parameter space~\cite{Liao.2010,devreese.klimin.11,wang.che.17}.
Another kind of pairing and phase coherence that can appear is the spatially homogeneous spin-polarized superfluidity (called breached pair state or Sarma phase \cite{sarma}), which has a gapless spectrum for the majority spin species.

Quite recently, the Rice University experimental group~\cite{Liao.2010} predicted the FFLO phase apperance in ultracold lattice gases.
The experimental setup allows to investigate imbalanced quantum Fermi gases ($N_{\downarrow}\neq N_{\uparrow}$) by trapping the two lowest hyperfine levels of the $^{6}$Li ground state in quasi-1D geometries~\cite{zhao.liu.09}. Similar experiments have been performed for the mass-imbalanced mixtures of $^{6}$Li and $^{40}$K atoms~\cite{wille.spiegelhalder.08,voigt.taglieber.09,trenkwalder.kohstall.11,jag.zaccanti.14}.

Also theoretical analyses suggest a possibility of a realization of the FFLO phase in optical lattices~\cite{rosenberg.chiesa.15,machida.mizushima.06,cichy.cichy.15}.
The existence of non-zero total momentum Cooper-pairs leads to a spontaneous symmetry breaking of the order parameter in real space~\cite{matsuda.shimahara.10}.
It is manifested by sign change of the superconducting order parameter as well as the occurrence of nodal lines in real space.
The spatial profile and the number of the nodal lines depend on the magnetic field ~\cite{shimahara.98}.
The same behaviour of the order parameter can be observed in ultracold fermionic gases in parabolic or toroidal traps.
In the former case there can occur oscillations of the order parameter in the radial direction~\cite{machida.mizushima.06,castorina.grasso.05,chen.wang.09}, whereas in the latter case the breaking of the rotational symmetry can result  in oscillations of the order parameter depending on the angle~\cite{yanase.09,ptok.12}.

For one-dimensional two-component Fermi atomic gases in a magnetic trap the exact thermodynamic Bethe ansatz~\cite{guan.batchelor.13} solution shows that (in some range of magnetic field and in the strong coupling limit) a mixed phase with the two-shell structure with a partially polarized  superfluid core surrounded (analogous to the FFLO phase) by either a fully paired or fully polarized phases occurs in the ground state~\cite{guan.batchelor.07,orso.07}.
Similar situation has been found in a case of the one-component trapped gas~\cite{hu.xiaji.07}.
Moreover, the FFLO phase occurs at all non-zero partially polarization for any attractive interaction, whereas all of the phase transitions are continuous~\cite{hu.xiaji.07,guan.batchelor.07,orso.07}.
Theoretical investigations predict that the FFLO state can be also realized in a case of the mass-imbalance fermionic system~\cite{liu.wilczek.03,pahl.koinov.14,
hu.maska.15,chung.bolech.16}.
There has been work on exact numerical studies (Quantum Monte Carlo (QMC) simulations and Density Matrix Renormalization Group (DMRG)) of the 1D attractive Hubbard model with population-imbalanced fermions~\cite{feiguin.heidrich.07,
batrouni.huntley.08,luscher.noack.08,rizzi.polini.08,tezuka.ueda.08,
burovski.orso.09,heidrichmeisner.orso.10,heidrichmeisner.feiguin.10,
riegger.orso.15,chung.bolech.16}, suggesting that the FFLO state is stable in 1D.
Indeed, the instability of the normal state with respect to FFLO is due to a Fermi surface ``nesting'' which is enhanced in 1D \cite{parish.2007}.

Motivated by the experimental feasibility of such systems with ultracold gases loaded on a quasi-1D lattice, we study the unconventional superfluid phases of the attractive Hubbard model (AHM) ($U<0$), in the presence of an external magnetic field.
We show that with increasing magnetic field, the system can evolve from the BCS-type superconducting state to the FFLO phase (where the Cooper pairs have non-zero total momentum ${\bm Q}$).
In an extremal case, this momentum ${\bm Q}$ can lie on the vertex of the first Brillouin zone (FBZ)~\cite{mierzejewski.maska.04,ptok.mierzejewski.08,ptok.maska.09} and the so-called $\eta$ phase emerges.
It should be stressed that the Hubbard model on a bipartite (alternate) lattice has been rigorously proved to have $\eta$ states as eigenstates~\cite{yang.89}.
Moreover, $\eta$-pairing has been found as a mechanism of superconductivity in a large class of models of strongly correlated electron systems (extended Hubbard models)~\cite{deboer.korepkin.95}.

We obtain the magnetic field vs. chemical potential as well as vs. filling (i.e $h-\mu$ and $h-n$, respectively) phase diagrams for several values of the on-site pairing interactions.
Therefore, the results of our analysis can be compared to experimental results where the filling or particle concentrations can be fully controled and measured.
We find a topological quantum phase transition, of the Lifshitz type, in the ground state phase diagrams.
A consequence of this transition is a change of the Fermi surface (FS) topology due to the variation of the Fermi energy and/or the band structure.

The paper is organized as follows.
In Section~\ref{sec.2}, we introduce the main theoretical model for the system under study, the attractive Hubbard model in a Zeeman magnetic field and we shortly discuss the mean-field method.
Section~\ref{sec.3} presents numerical results and their discussion: the $h-\mu$ as well as $h-n$ phase diagrams in the weak-coupling limit (\ref{subsec.3a}), the BCS-BEC crossover analysis and magnetic Lifshitz transition (\ref{subsec.3b}).
We conclude in Sec.~\ref{sec.4} with a brief summary of the obtained results and an outlook.

\section{Model and method}
\label{sec.2}

We study an $s$-wave superconductor on a one-dimensional lattice, described by AHM ($U<0$) in a magnetic field which in real space takes the form:
\begin{eqnarray}
\label{ham_real}
\mathcal{\hat H} = \sum_{ \langle i,j \rangle \sigma } \left( - t - (\mu + \sigma h ) \delta_{ij} \right) \hat c_{i\sigma}^{\dagger} \hat c_{j\sigma} + U \sum_{i} \hat n_{i\uparrow} \hat n_{i\downarrow},
\end{eqnarray}
where $t$ is the nearest-neighbor hopping, $\sigma=\uparrow ,\downarrow$ the spin index, $U$ the on-site attraction, $\mu$ is the chemical potential; $h$ is a Zeeman field, which originates from an external magnetic field (in $g \mu_B \slash 2$ units) or from a population imbalance in the context of the cold atomic Fermi gases with $\mu =(\mu_{\uparrow} +\mu_{\downarrow})/2$ and $h=(\mu_{\uparrow} - \mu_{\downarrow})/2$, where $\mu_{\sigma}$ is the chemical potential of atoms with (pseudo) spin-$\sigma$.
The second term can be decoupled using the mean-field approximation,
\begin{eqnarray}
\hat n_{i\uparrow} \hat n_{i\downarrow} = \Delta_{i}^{\ast} \hat c_{i\downarrow} \hat c_{i\uparrow} + \Delta_{i} \hat c_{i\uparrow}^{\dagger} \hat c_{i\downarrow}^{\dagger} - | \Delta_{i} |^{2},
\end{eqnarray}
where $\Delta_{i} = \langle \hat c_{i\downarrow} \hat c_{i\uparrow} \rangle$ is defined as the superconducting order parameter (SOP). Then, the mean-field Hamiltonian in real space takes the form
\begin{eqnarray}
 \mathcal{\hat H}^{MF} &=& \sum_{ \langle i,j \rangle \sigma } \left( - t - ( \mu + \sigma h ) \delta_{ij} \right) \hat c_{i\sigma}^{\dagger} \hat c_{j\sigma} \\
\nonumber &+& U \sum_{i} \left( \Delta_{i}^{\ast} \hat c_{i\downarrow} \hat c_{i\uparrow} + H.c. \right) - U \sum_{i} | \Delta_{i} |^{2}.
\end{eqnarray}
Without loss of generality, we can write down the SOP as: $\Delta_{i} = \Delta_{0} \exp ( i {\bm Q} \cdot {\bm R}_{i} )$, where $\Delta_0$ is the spatially oscillating amplitude, while ${\bm Q}$ is the total momentum of the Cooper pair.

Transforming the Hamiltonian (\ref{ham_real}) to the reciprocal space, one obtains
\begin{eqnarray}
\label{ham_recip}
\mathcal{\hat H}^{MF} &=& \sum_{ {\bm k} \sigma } E_{{\bm k}\sigma} \hat c_{{\bm k}\sigma}^{\dagger} \hat c_{{\bm k}\sigma} \\
\nonumber &+& U \sum_{\bm k} \left( \Delta_{0}^{\ast} \hat c_{-{\bm k}+{\bm Q}\downarrow} \hat c_{{\bm k}\uparrow} + \text{H.c.} \right) - U N | \Delta_{0} |^{2}.
\end{eqnarray}
In the one-dimensional lattice case, the dispersion relation is given by: $E_{{\bm k}\sigma} = - 2 t \cos( k_{x} ) - \left( \mu + \sigma h \right)$. Using the Nambu notation, the Hamiltonian (\ref{ham_recip}) can be rewritten in a matrix form, $\mathcal{\hat H}^{MF} = \sum_{\bm k} \hat\Phi_{\bm k}^{\dagger} \mathbb{H}_{\bm k} \hat\Phi_{\bm k}$, with
\begin{eqnarray}
\mathbb{H}_{\bm k} = \left(
\begin{array}{cc}
E_{{\bm k}\uparrow} & U \Delta_{0} \\
U \Delta_{0}^{\ast} & -E_{-{\bm k}+{\bm Q}\downarrow}
\end{array}
\right),
\end{eqnarray}
where $\hat\Phi_{\bm k}^{\dagger} = ( \hat c_{{\bm k}\uparrow}^{\dagger} , \hat c_{-{\bm k}+{\bm Q}\downarrow} )$ are the Nambu spinors. Then, the eigenvalues $\lambda_{{\bm k}\pm}$ of $\mathcal{\hat H}^{MF}$ are given by
\begin{eqnarray}
\lambda_{{\bm k}\pm} &=& \eta^{-}_{\bm k} \pm \vartheta_{\bm k}, \\
\nonumber \eta^{\pm}_{\bm k} = \frac{ E_{{\bm k}\uparrow} \pm E_{-{\bm k}+{\bm Q}\downarrow} }{2} , &\quad& \vartheta_{\bm k} = \sqrt{ \left( \eta^{+}_{\bm k} \right) ^{2} + U^{2} | \Delta_{0} |^{2} } .
\end{eqnarray}
The grand canonical potential defined by $\Omega \equiv - k_{B} T \ln \{ \Tr [ \exp ( - \mathcal{\hat H}^{MF} / k_{B} T ) ] \}$ can be written as
\begin{eqnarray}
\label{gcp}
\Omega &=& - k_{B} T \sum_{{\bm k},\alpha\in\pm} \ln \left( 1 + \exp \left( \frac{ - \lambda_{{\bm k}\alpha} }{ k_{B} T } \right) \right) \\
\nonumber &+& \sum_{\bm k} \left( E_{{\bm k}\downarrow} - U | \Delta_{0} |^{2} \right) ,
\end{eqnarray}
while the particle number equation takes the form
\begin{eqnarray}
n \equiv \frac{ - 1 }{N} \frac{ \partial\Omega }{ \partial\mu } = 1 + \frac{1}{N} \sum_{\bm k} \frac{ \eta^{+}_{\bm k} }{ \vartheta_{\bm k} } \left( f ( \lambda_{{\bm k},+} ) - f ( \lambda_{{\bm k},-} )  \right) .
\end{eqnarray}

The ground state is found by a minimization of $\Omega$ with respect to the SOP amplitude $\Delta_{0}$ and momentum ${\bm Q}$, for fixed $\mu$ and $h$, at a temperature $T/t= 10^{-5}$ (effectively $T=0$, non-zero value taken for numerical reasons). As mentioned above, the systems in which the FFLO phase can be realized are characterized by discontinuous phase transitions, which are associated with discontinous changes of $\Delta_{0}$ and/or ${\bm Q}$. As a consequence, the energy gap equation for a given phase, equivalent to one of the conditions of the energy minimization $d\Omega/d\Delta_{0} = 0$, at fixed ${\bm Q}$, cannot be used for the phase boundaries estimation.
In this case, the procedure of the minimization of $\Omega$ with respect to the SOP amplitude and all possible momenta ${\bm Q}$ realized in the system is essential.
Because of the unequivocal relation of the real space and reciprocal space (via the Fourier transform), the number of possible ${\bm Q}$ vectors in the lattice is equal to the number of lattice sites (given by $N$).
It is worth to mention that ${\bm Q}$ as well as $\Delta_0$ change discontinuously~\cite{ptok.15} going from the BCS to the FFLO phase.
To find the minimum of the energy of the system, one minimizes $\Omega ( \Delta_{0} )$ functions for $N$ different ${\bm Q}$ vectors.
For simplicity, without loss of generality, numerical calculations have been performed in the lattice with $N = 200$ sites and periodic boundary conditions, which makes the finite-size effects negligible~\cite{ptok.crivelli.16}.
To speed up the calculations, graphical cards have been used. We have proceeded according to the numerical procedure described in Ref.~\cite{januszewski.ptok.15}.

\section{Numerical results and discussion}
\label{sec.3}

In this section, we focus on the analysis of superconducting properties of ultracold atomic mixtures assuming a one-dimensional lattice geometry. Within the mean-field (BCS-Stoner)
approach, we construct the phase diagrams in two ways: by fixing the chemical
potential ($\mu$) or the particle concentration ($n$), and show the relevant
differences resulting from these possibilities. The ground-state phase diagrams are obtained for a wide range of attractive interactions, i.e. for a weak and intermediate coupling (\ref{subsec.3a}) and for the local pairs limit (BEC) (\ref{subsec.3b}) by using the mean field approximation.
Notice that in general case this approximation overestimates critical temperatures and can give an incorrect description of the long-range order phases.
However, it gives a relatively good description of the system in the ground state (at $T=0$), even in the strong coupling limit~\cite{micnas.ranninger.90}.

\subsection{Superconducting properties of the AHM in the presence of a Zeeman magnetic field: Weak and intermediate coupling}
\label{subsec.3a}

\begin{figure}[!t]
\centering
\includegraphics[width=\linewidth]{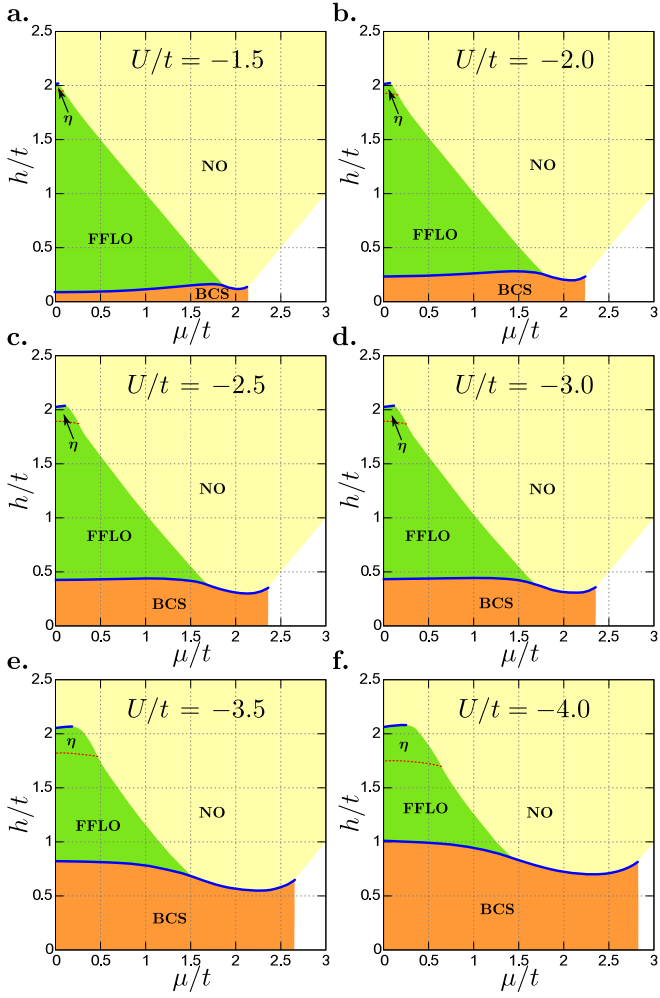}
\caption{(Color online)
	$h-\mu$ ground state phase diagram for several values of the pairing interaction $U$.
	Labels are as follows: NO -- normal phase, BCS -- non-polarized superconducting state with ${\bf Q}=0$, FFLO -- polarized superconducting phase with ${\bf Q}\neq 0$.
	Additionally, within the FFLO phase, above the dashed red line there is a region where the $\eta$ phase is distinguished.
	The solid blue lines indicate first order phase transitions between different states. The white region indicates the empty state (or fully-filled state, depending on the sign of the chemical potential).
\label{fig.df.mu}
}
\end{figure}

\begin{figure}[!t]
\centering
\includegraphics[width=\linewidth]{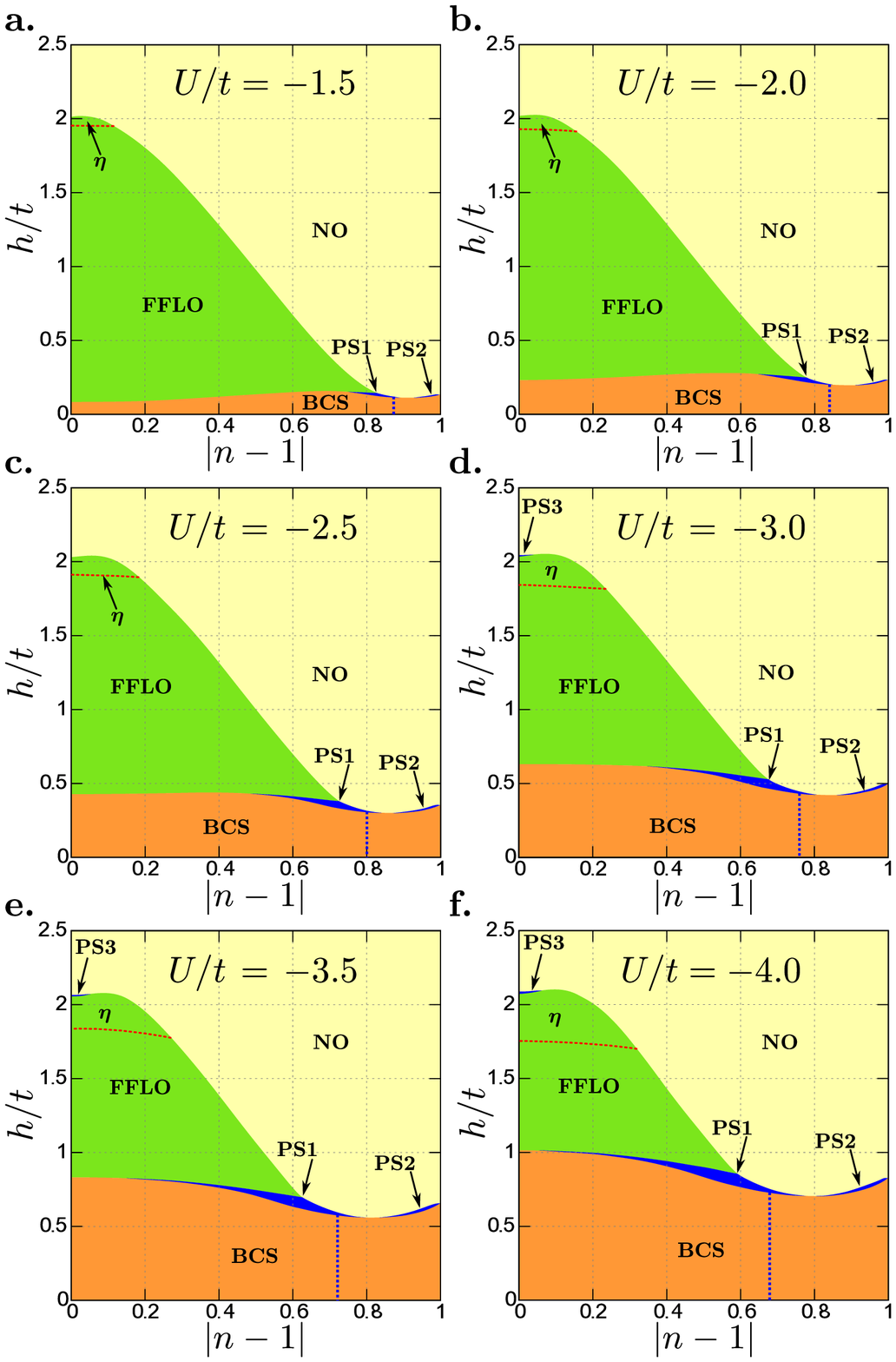}
\caption{(Color online)
	$h-n$ ground state phase diagram for several values of the pairing interaction $U$.
	Labels are as follows: NO -- normal phase, BCS and FFLO -- superconducting states with ${\bf Q}=0$ and ${\bf Q}\neq 0$, respectively,
	PS1 -- phase separation region between BCS and FFLO or between BCS and NO states (details in the text),
	PS2 -- phase separation region between BCS and NO phases,
	PS3 -- phase separation region between FFLO ($\eta$-pairing) and NO states.
	Within the FFLO phase the $\eta$ state exists above the dashed red line.
\label{fig.df.n}
}
\end{figure}

In this subsection, we consider the ground state phase diagrams in the weak and intermediate couplings.
In the following, we set $t=1$.

In the weak coupling regime and in absence of an external Zeeman field, the usual superconducting BCS-type s-wave state is stable (Fig. \ref{fig.df.mu}). As the magnetic field rises, superfluidity gets destroyed, at weak and intermediate couplings, due to paramagnetic effects or by population imbalance. Hence, the unpolarized BCS-like superconducting phase undergoes a first order phase transition to the polarized normal state or to the FFLO phase. Rising higher the field and close to half-filling, the polarized FFLO-$\eta$-pairing superconducting phase also undergoes a first order phase transition to the normal state.

These two first order phase transition lines were determined from the conditions: $\Omega^\text{BCS}=\Omega^\text{FFLO}$, $\Omega^\text{BCS}=\Omega^\text{NO}$, $\Omega^\text{FFLO}=\Omega^\text{NO}$, where $\Omega^\text{BCS}$, $\Omega^\text{FFLO}$ and
$\Omega^\text{NO}$ denote the grand canonical potential of the BCS ($\Delta_0 \neq 0$, ${\bf Q}=0$, $P= 0$), FFLO ($\Delta_0 \neq 0$, ${\bf Q}\neq 0$, $P\neq 0$) and the normal ($\Delta_0= 0$, $P\neq 0$) state,
respectively, where $P=(n_{\uparrow}-n_{\downarrow})/(n_{\uparrow}+n_{\downarrow})$ is the polarization.
Then, these results have been mapped onto the case of fixed $n$ (Fig.~\ref{fig.df.n}). Moreover, there is a special case of the FFLO phase, for which the Cooper pair momentum takes the value of the momentum on the FBZ vertex ($|{\bf Q}|= \pi$). This case is called $\eta$-pairing and is found in the phase diagrams as well. It is worth to mention that we take into account in our analysis the Sarma phase, which is characterized by the spatially homogeneous order parameter, in the presence of non-zero polarization (i.e., $\Delta_0 \neq 0$, ${\bf Q}=0$, $P\neq 0$). However, these solutions are unstable for the whole region of parameters.

\paragraph*{$h$ vs. $\mu$ phase diagram.}
Fig. \ref{fig.df.mu} shows the $h-\mu$ phase diagrams at $T=0$.
These diagrams are symmetric with respect to the sign change of $\mu$ or $h$ due to the particle-hole symmetry. For the sake of clarity, we only show the range of $\mu$ from 0 to 3 and $h\geq 0$.
In this case (see Fig.~\ref{fig.df.mu}), we find two types of superconducting phases: BCS and FFLO type. Note that inside the latter, above some magnetic field value identified by a red dashed line, we find the $\eta$-FFLO phase.
The stability range of the BCS state as well as the $\eta$-pairing depends on the value of the attractive interaction -- both phases widen when increasing the attraction and the FFLO phase shrinks.
	Notice that obtained phase diagrams (Fig.~\ref{fig.df.mu}) are in a qualitative agreement with the previous DMRG calculation performed for trapped spin-imbalanced Fermi gas, where partially polarized state
	 (i.e. the FFLO state in the present paper) exists in a large range of the model parameters~\cite{heidrichmeisner.orso.10}.

\paragraph*{Phase transitions.}
We find that the phase transition from the BCS phase to the FFLO or NO state is always of the first order (associated with a discontinuous change of the order parameters).
At relatively high magnetic field and around half-filling ($\mu \simeq 0$), we also obtain a first order phase transition from the $\eta$-FFLO phase to the NO state, (blue solid lines in Fig.~\ref{fig.df.mu}),
whereas the transition for larger $\mu$ changes its nature into second order.
On the other hand, 
the transitions from the FFLO phase to the NO state as well as between the BCS phase and the empty  (full-filled) state are second order ones (connected with a continuous change of the order parameters).
It is important to emphasize that the first order phase transitions are reflected by the existence of the phase separation (PS) regions in the $h-n$ phase diagrams.
The BCS boundary shows strong non-linearities, especially around the BCS-BEC crossover point (for $|\mu| \simeq 2$), while the boundary between the FFLO phase and the NO state changes in an approximately linear way with $\mu$.

Generally, the order of the phase transition between the FFLO and BCS phases is still  under debate~\cite{casalbuoni.nardulli.04,dalidovich.yang.04,samokhin.marenko.06,konschelle.cayssol.07,stoof.gubbels.09,caldas.continentino.12,buchhold.everest.17}.
For 1D systems the studies of that problem within the framework of the Ginzburg-Landau theory show that e.g. disorder can change the type of the phase transition~\cite{agterberg.yang.01}.
A combination of the renormalization group and mean-field approximation for  Fermi gases with attractive interaction gives second-order phase transition between uniform (BCS) and nonuniform (FFLO) superconducting states~\cite{yang.01}.
Moreover,  studies of two-component Fermi atomic gases in a magnetic trap using the exact thermodynamic Bethe ansatz solution in continuum model show that the all of the phase transitions are continuous~\cite{hu.xiaji.07,guan.batchelor.07,orso.07}.
In such systems there has been shown that the phase separation in a real space can occur which can be source of other types of the phase transitions.
The effective Ginsburg-Landau theory studies for quasi-2D {\it d}-wave superconductors by renormalization group analysis indicate that the transition form the FFLO to normal state is generically first order, even when the mean-field theory suggests a continuous transition~\cite{dalidovich.yang.04}.

\paragraph*{$h$ vs. $n$ phase diagram and phase separations.}
As mentioned above, there are relevant differences between the phase diagrams obtained for fixed chemical potential and fixed particle concentration.
Fig. \ref{fig.df.n} shows the dependence of the critical magnetic fields on the filling, for several attraction values.
Here, due to the particle-hole symmetry, we only show the range of $|n-1|$ from 0 to~1.
In contrast to the fixed chemical potential case,
if the number of particles is fixed and $n\neq 1$, the phase separated states are present on the diagrams.
The occurrence of the phase separated states for fixed concentration (so-called macroscopic phase separation) is associated with the first order phase transitions occurring for fixed $\mu$~\cite{kapcia.robaszkiewicz.12,kapcia.czart.16}.
Due to the fact that the transition for fixed $\mu$ between the FFLO and BCS  phases, the BCS and NO phases, and the $\eta$-FFLO and NO phases (in some ranges of the model parameters) are discontinuous, the corresponding phase separated states are present on the diagrams as a function of $n$.
One can distinguish three different phase separation regions in the $h-|n-1|$ phase diagram:
PS1 -- the region of phase separation between BCS and FFLO phases as well as between the BCS and NO phases (for $h$ above and below, respectively, the points indicated by the arrow in Fig.~\ref{fig.df.n}),
PS2 -- between the BCS and NO phases, and
PS3 -- between the $\eta$-FFLO-pairing and NO states.

The phase diagrams in Fig.~\ref{fig.df.n} show that the FFLO phase can be realized at relatively large doping.
Similarly as in the case of the $h$-$\mu$ phase diagram, the phase boundaries show strong non-linearities in the BCS-BEC regime (small density of particles). The blue dashed vertical line indicates the critical value of $n$ above which, according to the Leggett criterion, there is the BCS-BEC crossover at the large spin-imbalance or Zeeman fields.

\begin{figure}[!b]
\centering
\includegraphics[width=\linewidth]{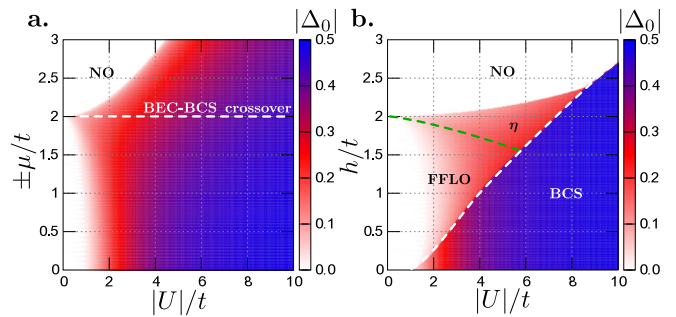}
\caption{(Color online)
	The influence of the pairing $U$ interaction  on the BCS-BEC region for $h = 0$ (a) and the magnetic Lifshitz transition region for $\mu = 0$ (b); color-coded -- $\Delta_0$ -- the amplitude of the order parameter.
\label{fig.becbcs2}
}
\end{figure}

\paragraph*{Role of the pairing $U$ interaction.}
The increasing of the pairing interaction $U$ leads to the stabilization of some critical behaviours (Fig.~\ref{fig.becbcs2}).
As it is known, in the strong coupling limit of AHM ($h=0$), the tightly bound local pairs of fermions behave as hard-core bosons and can exhibit a superfluid state similar to that of
$^{4}\textrm{He}$~II \cite{micnas.ranninger.90}.
According to the Leggett criterion, the Bose regime begins when the chemical potential $\mu$ drops below the lower band edge.
In the case of a one-dimensional system, the band edges are at $\pm \mu/t = 2$. Fig.~\ref{fig.becbcs2}(a) shows the $\mu$ vs. $U$ phase diagram, at $T=0$ and $h=0$.
As one can notice, in the case of the strong coupling (larger values of $U$), the superconducting phase exists above the band boundary (white dashed line).
Above this line, one can speak about non-BCS behaviour.

We observe similar critical behaviour with an increasing Zeeman magnetic field (Fig.~\ref{fig.becbcs2}(b)).
Namely, at the critical point, $U=0$ and $h/t = 2$, the magnetic Lifshitz transition (MLT)~\cite{ptok.kapcia.17} takes place.
As has been mentioned above, at a non-zero Zeeman magnetic field, the population imbalance introduces a mismatch between the Fermi surfaces.
Hence, effectively there are two Fermi surfaces in the system, one for the majority spin component and one for the minority spin component. However, above $h/t = 2$ (the value of the band edge), one of the Fermi surfaces disappears. Therefore, one can observe a change in the FS topology.
Strikingly, the superconducting phase can still survive above MLT and the increasing of $U$ stabilizes the $\eta$ phase (Fig.~\ref{fig.becbcs2}(b)).
The boundary between the FFLO and $\eta$ phase is moved towards lower values of the magnetic field, which is clearly visible in the $h-\mu$ as well as $h-|n-1|$ phase diagrams.

\subsection{BCS-BEC crossover and magnetic Lifshitz transition}
\label{subsec.3b}

\begin{figure}[!b]
\centering
	\includegraphics[width=\linewidth]{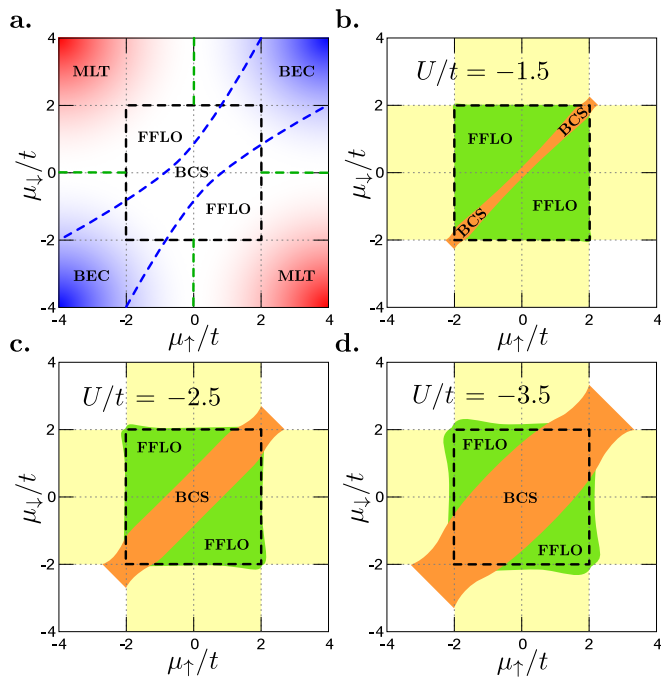}
\caption{(Color online)
	Ground state effective chemical potentials $\mu_{\uparrow} = \mu+h$ vs. $\mu_{\downarrow} = \mu-h$ phase diagram.
	(a) A schematic diagram. The dashed black square indicates the band edges. The inside of the square shows possible regions of the BCS, FFLO, BEC state. The red region -- magnetic Lifshitz transition (MLT) occurrence, the blue region -- the BCS-BEC crossover region. The dashed blue lines show the schematic boundaries of the BCS and FFLO phases occurrence.
	(b)--(d) Results for different values of the paring interaction $U$.
\label{fig.becbcs}
}
\end{figure}

In this subsection, we present results concerning the BCS-BEC
crossover as well as the magnetic Lifshitz transition.
Both possibilities can be simply shown by means of $\mu_{\uparrow}$ vs. $\mu_{\downarrow}$ phase diagrams (Fig.~\ref{fig.becbcs}), where $\mu_{\sigma} = \mu + \sigma h$ is the effective chemical potential.

First, let us discuss the schematic phase diagram in Fig.~\ref{fig.becbcs}(a).
In the weak coupling limit, for
$|\mu_{\sigma}|/t \leq 2$ (the inside of the dashed black square), we have the BCS phase or FFLO, depending on the population imbalance. If $\mu_{\uparrow}\simeq\mu_{\downarrow}$ ($i$), there is the unpolarized BCS phase, otherwise (for $\mu_{\uparrow}\neq \mu_{\downarrow}$ ($ii$)) the FFLO state is stable. In Fig.~\ref{fig.becbcs}(a), dashed blue lines indicate the schematic boundaries between the BCS phase and the FFLO state. In Fig.~\ref{fig.becbcs}(b)-(d), these boundaries are obtained from the minimization of the grand canonical potential with respect to the amplitude of the order prameter $\Delta_0$ and the vector ${\bf Q}$.

With increasing $U$, when the effective chemical potentials  drop below the lower band edge, there is the crossover to the tightly bound local pairs region (BEC -- the blue shaded area in Fig.~\ref{fig.becbcs}(a)). It takes place in the region of parameters for which $\mu_{\uparrow}\simeq\mu_{\downarrow}$, i.e. the polarization $P$ of the system is low. However, it is worth to emphasize that the BCS-BEC crossover takes place both for a low particle concentration ($\mu=(\mu_{\uparrow}+\mu_{\downarrow})/2<-2t$) and for a low concentration of holes ($\mu=(\mu_{\uparrow}+\mu_{\downarrow})/2>2t$). This behaviour is clearly visible in Figs.~\ref{fig.becbcs}(b)-(d) in which the orange area significantly exceeds the black dashed square. Therefore, one can speak of the crossover to the tightly bound local pairs.

\begin{figure}[!t]
\centering
\includegraphics{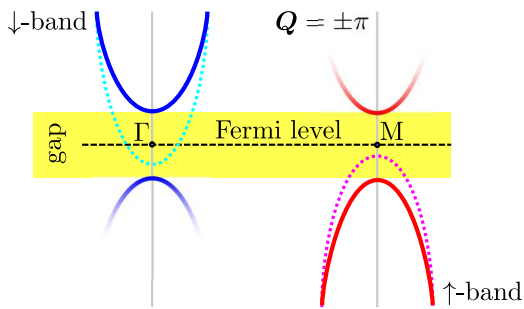}
\caption{(Color online)
A schematic illustration of the (quasi-)particle band structure above the magnetic Lifshitz transition. The solid (dashed) lines indicate the quasi-particle (particle) bands in the superconducting (normal) state. At relatively high magnetic field and interaction, the eta-pairing can be realized (which is the superconducting state with total momentum of the Cooper pairs equal $M$ point of the first Brillouin zone). In this case, the magnetic field causes a splitting of bands with opposite spins and hence, one of the bands can be fully filled (empty), whereas the top/bottom of the other band crosses the Fermi level. However, the existence of a strong pairing interaction (and the energy gap) leads to pairing and modifies the quasi-particle spectrum in a similar manner to the one in the BCS-BEC crossover regime~\cite{cuoco.ranninger.06,lubashevsky.2012,loh.randeria.16}, at ${\bm Q} > 0$.
\label{fig.schem}
}
\end{figure}

In the weak coupling region, in the presence of a Zeeman magnetic field, there are two FS's in the system.
If the system is strongly-polarized, i.e. $\mu_{\uparrow}+\mu_{\downarrow}\sim 0$, the magnetic Lifshitz transition can take place.
In this case, one of the spin bands is fully-filled or empty. However, the increase of the attractive interaction leads to the stabilization of the superconducting state, although there is only one FS in the system (see: Fig.~\ref{fig.schem}).
There is the pairing between the particles with opposite spins but the total momentum of pairs (${\bf Q}$) takes the maximum allowed value of momentum in the system (i.e. the vertex of FBZ), in the presence of high polarization.

The phase diagrams for different values of pairing interaction $U$ are shown in panels b-d of Fig.~\ref{fig.becbcs}.
As is clearly visible, in the weak coupling limit ($| U/t | \rightarrow 0$), the BCS phase ($\mu_{\uparrow}\simeq\mu_{\downarrow}$) as well as the FFLO state ($\mu_{\uparrow}\neq \mu_{\downarrow}$) are stable.
However, the increasing $U$ widens the range of occurrence of the BEC and MLT regions, which is clearly visible in panel d. In the strong coupling limit, the chemical potential drops below the band edge, the Fermi surfaces disappear and the FFLO phase is unstable. In this regime, only the unpolarized superconducting state is realized.

\section{summary}
\label{sec.4}

We studied the superconducting properties of the spin-imbalanced attractive Hubbard model in the context of experiments with ultracold atomic Fermi mixtures with population imbalance in one-dimensional optical lattices. The ground-state phase diagrams were obtained for the cases of a fixed chemical potential and a fixed density (lattice filling) by using the mean-field approach for the 1D system.
We found that the FFLO phase is stabilized for a wide range of atomic densities due to a Fermi
surface ‘‘nesting’’, which is enhanced in 1D. Superconductivity is destroyed by the pair breaking in a very weak coupling
regime. If the number of particles is fixed and $n\neq 1$, one can obtain two critical Zeeman
magnetic fields (population imbalance), which limit the phase separation of the superconducting
and the normal states.

At relatively high values of a Zeeman magnetic field, there is a region of the $\eta$-pairing (within the FFLO phase).
With an increasing attractive interaction, the $\eta$-pairing is stabilized with respect to the FFLO state. Moreover, the $\eta$ phase can be stable even above the magnetic Lifshitz transition (Fig.~\ref{fig.schem}). A consequence of this transition is a change of the Fermi surface topology due to the variation of the Fermi energy and/or the band structure. Our finding of a MLT in the spin-imbalanced AHM in one-dimensional lattice and determination the stability of $\eta$-pairing is reported for the first time in the literature.

Hence, at $T=0$, in the weak coupling regime and for fixed $n$, the following states have been found in the 1D system: at $h\geq 0$ -- the BCS state; for higher values of magnetic fields ($h\neq 0$) -- the FFLO phase;  at relatively high $h\neq 0$ -- the $\eta$-pairing; three different PS regions and NO. PS terminates at tricritical points.

We have also investigated the ground state BCS-BEC crossover diagrams for AHM in
the presence of a Zeeman magnetic field. We have observed that the FFLO phase is suppressed with increasing attraction, the $\eta$-pairing is favoured as well as only the BCS-like phase in the strong coupling limit.

\begin{acknowledgments}
We thank Krzysztof Cichy, Tadeusz Doma\'{n}ski and Matteo Rizzi for careful reading of the manuscript, valuable comments and discussions.
This work was supported by the National Science Center (NCN, Poland) under grant UMO-2016/20/S/ST3/00274 (A.P.) and by the Large Infrastructures for Research, Experimental Development and Innovations project "IT4Innovations National Supercomputing Center --- LM2015070" of the Czech Republic Ministry of Education, Youth and Sports.
K.R. acknowledges the support from CIBioFi and the Colombian Science, Technology and Innovation Fundation--COLCIENCIAS ``Francisco Jos\'e de Caldas'' under project 1106-712-49884 (contract No.264-2016) and --General Royalties System (Fondo CTeI-SGR) under contract No. BPIN 2013000100007.
\end{acknowledgments}

\bibliography{biblio}

\end{document}